\documentclass[11pt]{article}
\usepackage{blois,epsfig,graphicx}

\def\p{\phi}
\def\eps{\epsilon}

\def\al{\alpha}

\def\be{\begin{equation}}
\def\ee{\end{equation}}
\def\bea{\begin{eqnarray}}
\def\eea{\end{eqnarray}}

\begin{document}
\vspace*{4cm}
\title{OBSERVABLE GRAVITATIONAL WAVES AND SPECTRAL INDEX RUNNING IN SMALL SINGLE FIELD INFLATIONARY MODELS\footnote{This is a contribution to the proceedings of "Windows to the Universe - Blois 2009". A more elaborate discussion can be found in \cite{BenDayan:2009kv}.}}

\author{ IDO BEN-DAYAN }

\address{Department of Physics, Ben-Gurion University,
    Beer-Sheva 84105, Israel}

\maketitle\abstracts{
 I construct a class of single small field models of inflation that can predict an observable gravitational wave signal in the cosmic microwave background anisotropoies, contrary to popular wisdom. The spectral index, its running, the tensor to scalar ratio and the number of e-folds can cover all the parameter space currently allowed by cosmological observations. A unique feature of models in this class is  their ability to predict a negative spectral index running in accordance with recent cosmic microwave background observations. I comment on the new class of models from an effective field theory perspective and show that if the dimensionless trilinear coupling is small, as required for consistency, then the observed spectral index running implies a high scale of inflation and hence an observable gravitational wave signal.
}

\section{The Model, Notations and Observables}
Consider a single canonically normalized scalar field slowly rolling down the potential for 60 e-folds. The potential comes from some underlying particle theory and is Taylor expanded around the CMB point. The potential and number of e-folds are:
\bea
\label{poten}
\frac{V(\p)}{V(0)}&=&1 - \sqrt{\frac{r_0}{8}}\p + \frac{\eta_0}{2} \p^2 - \frac{\al_0}{
 3 \sqrt{2r_0}}\p^3 - a_4 \p^4 - a_5 \p^5\\
N_{CMB} &=& \int_0^{\p_{END}}\frac{d\p}{\sqrt{2 \eps(\p)}}
\eea
where the field distance is measured in the reduced Planck mass $M_p\equiv 1$ and all couplings are dimensionless. Lets use the regular slow-roll parameters, $
\epsilon=\frac{1}{2}\left(\frac{V'}{V}\right)^2
$,
$
\eta=\frac{V''}{V}
$, and
$
\xi^2=\frac{V^{'''}V'}{V^2}
$.
Without loss of generality, the CMB point is taken at the origin. If the running $\al$ is large enough it affects the spectral index which is calculated as follows: 
\be
n_s=1+2\eta-6\eps+2\left[\frac{1}{3}\eta^2+(8C-1)\eps \eta-\left(\frac{5}{3}-C\right)\eps^2-\left(C-\frac{1}{3}\right)\xi^2\right]\\
\ee
where $C=-2+\ln 2+\gamma$, $\gamma$ being the Euler constant.
The tensor to scalar ratio and running are unaffected at this level and are given by the "usual"
$r = 16 \epsilon$ and
$
\alpha = -16 \epsilon \eta+24\epsilon^2+2\xi^2.
$
$r_0 ,\eta_0$ and $\al_0$ are the desired CMB observables. Parameters $a_4$ and $a_5$ determine the end of inflation and the number of e-folds.
\subsection{Small Field Models and Lyth Theorem}
The difference between small versus large field models is the distance the field traverses during the $60$ e-folds of observable inflation. In small field models the field rolls a distance of about $1 M_p$, while in large field models the distance is usually more than $10M_p$ . Small field models have several virtues: They are usually simple. No functional tuning is needed. They are easier to accommodate in string theory and do not depend sharply on initial conditions.

According to Lyth's theorem, $r\simeq 8(\Delta \phi/N_{CMB})^2<10^{-2}$ for small field models, thus rendering $r$ undetectable.
 I show that, contrary to popular wisdom\cite{lyththrm,BenDayan:2008dv}, interesting small field models can predict observable GW and observable negative running. These predictions are related to the fact that the rate of change of the Hubble parameter during the era when most e-folds were accumulated can be smaller than its value at the CMB point, thus evading Lyth's theorem.
\begin{figure}[t]
\hspace{1in}\psfig{figure=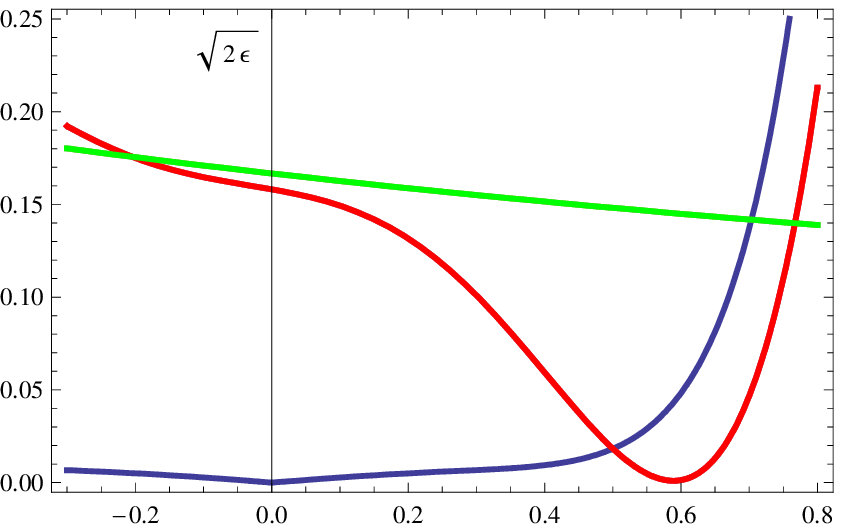,height=1in}\hspace{.5in}\psfig{figure=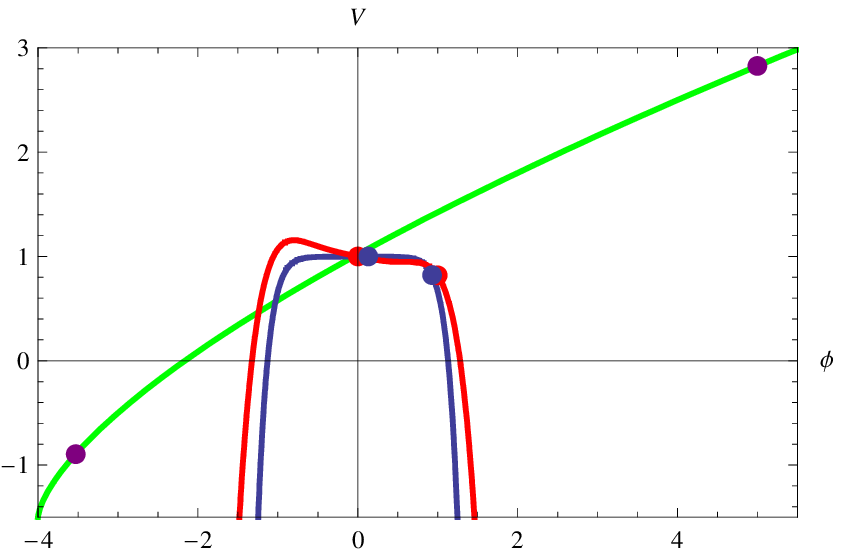,height=1.12in}
\caption{\label{figure1} A graph of $\sqrt{2 \epsilon}=|V'/V|$ (left) and $V$ (right)  for a small field canonical SUGRA model (blue), a large field model (green) and a model of the new class with non-monotonic $\epsilon$ (red). The new model interpolates between the two others. $\phi_{CMB}$ and $\phi_{END}$ are marked by the filled circles on the right panel. 
 The large field model is offset $V\rightarrow V-1.5$. Additionally,
to demonstrate the similarity between the small field model (blue) and the new model (red), a symmetric example is shown, i.e. $a_5=0, a_6=0.3911$. The CMB observables are $n_s=1.03, r=0.2, \al=-0.07$.
}
\end{figure}

The theorem assumes that $\epsilon$ is monotonic. This is not necessary. As seen in Figure 1, allowing $\epsilon$ to vary enables acquiring many e-folds away from the CMB point, thus yielding a large r at the CMB point while traveling only a short distance in field space. Usually, this causes other observables such as running to be large, or give insufficient number of e-folds and the higher order parameters constrain them to fit observations. This also explains why in large field models the field travels $10M_p$ to achieve detectable $r$, while from the theorem $2-3M_p$ seem enough. The new class of models "interpolates" between large and small field behavior as seen in Figure $1$ and produces a variety of observables as demonstrated in Table $1$ and Figure $2$.


\section{Results and Discussion}

\begin{table}[t]
\caption{The values of the potential parameters, the range of inflaton motion after $50$ and $60$ e-folds and the  values of the  CMB observables assuming that  $N_{CMB}=60$. The models appearing in Fig.~\ref{figure2}  are marked with an asterisk. The last two models are renormalizable models with $a_5=0$.}
\label{table:models}
\begin{center}
 \hbox{\hspace{1.2in}Potential parameters\hspace{1.1in}Range\hspace{.3in} CMB observables}\vspace{.05in}
\begin{tabular}{|c|r|r|r|l||c|c||c|c|r|}
  \hline
   \ $r_0$ & $\eta_0$ & $\al_0$ & $a_4$ &  $a_5$ & $\ \Delta\phi_{50}\ $ & $\ \Delta\phi_{60}\ $ & $n_s$ & $r$ & $\al$  \\
    \hline
   \ $*\ 0.05$ & $ -0.02 $ & $-0.001$ & $-0.1752$ & $ 0.1314 $ & $0.855$ & $1.5$ & $0.94$ & $0.05$ & $0.0002$  \\
   \hline
\ $ *\ 0.10$ & $ 0.015 $ & $-0.03$ & $-0.6102$ & $ 0.709 $ & $0.567$ & $1.0$ & $0.96$ & $0.10$ & $-0.031$  \\
   \hline
   \ $*\ 0.04$ & $ 0.07 $ & $-0.05$ & $-0.2739$ & $ 0.48 $ & $0.5$ & $1.0$ & $1.07$ & $0.04$ & $-0.052$  \\
   \hline
   \ $*\ 0.04$ & $ 0.025 $ & $ -0.02 $ & $-0.436$ & $ 0.574 $ & $ 0.525$ & $1.0$ & $1.01$ & $0.04$ & $-0.021$  \\
   \hline
   \ $*\ 0.13$ & $ 0.01 $ & $ 0.001 $ & $-0.4072$ & $ 0.367 $ & $ 0.705$ & $1.2$ & $0.97$ & $0.13$ & $0.001$   \\
   \hline
\ $*\ 0.05$ & $ 0.02 $ & $ -0.05 $ & $-0.425$ & $ 0.591 $ & $ 0.53$ & $1.0$ & $0.97$ & $0.05$ & $-0.051$   \\
\hline
\ $*\ 0.02$ & $ 0.015 $ & $ -0.04 $ & $-0.691$ & $ 1.33 $ & $ 0.39$ & $0.8$ & $0.98$ & $0.02$ & $-0.04$   \\
   \hline\hline
\ $\ 0.02 $ & $ 0.1144 $ & $ 0 $ & $0.0325$ & $\ \  0 $ & $ 0.8$ & $2$ & $1.23$ & $0.02$ & $-0.0022$   \\
   \hline
\ $*\ 0.01 $ & $ 0.065 $ & $ -0.133  $ & $0.671$ & $\ \ 0 $ & $ 0.315$ & $0.9$ & $0.99$ & $0.01$ & $-0.134$   \\
   \hline
   \end{tabular}
\end{center}
\end{table}

As can be seen from Table $1$ and the Figure $2$, all parameter space allowed by WMAP5+QUaD \cite{WMAP5,quad} is covered by the new class of models. The ability to incorporate running of the spectral index is unique. It is also clear from Figure $2$ that the running $\al$ is a better discriminator of models, than $r$ or $n_s$. This is important since the QUaD experiment reported a $2\sigma$ detection of negative spectral index running with central value \cite{quad} $\al=-0.05$, as can be seen from Figure $2$.

Moreover, applying effective field theory (EFT) techniques to assess the theoretical validity of models yields a non-trivial connection between $\al$ and $r$ and hence the scale of inflation. Consider a potential $V(\phi)=\Lambda^4(1+\sum_n \lambda_n(\phi/M_p)^n)$. In our notations, for example $\lambda_2=\eta_0/2$ etc. As analyzed in \cite{Burgess:2009ea}, $\lambda_3 \ll 1$ for a consistent EFT. This gives the following connection between $\al$ and $r$: $\al\simeq 2 \xi^2 =2 \frac{V''' V'}{V^2}$. Since $\frac{V'''}{V}=3!\ \lambda_3$ and since $\frac{V'}{V}=\sqrt{r/8}$ one obtains $r=2(\al/(3! \lambda_3))^2$.
Let us now define $r_{0.01}\equiv r/0.01$ and $\al_{0.05}\equiv |\al|/0.05$ and $\widehat\lambda_3\equiv 3! \lambda_3$. Then $r_{0.01}=.5\ \al_{0.05}^2 {\widehat\lambda_3}^{-2}$. Imposing the condition that the validity of the EFT implies $\lambda_3\ll 1$, leads to a lower bound on the GW strength
$
r_{0.01}>.5\ \al_{0.05}^2
$,
which implies that if the value of the running is the one observed by QUaD then one should expect an observable $r$.
Similarly, using the standard estimate $\Lambda \simeq 1\times 10^{16}\textrm{GeV} (r_{0.01})^{1/4}$ one obtains $\Lambda \simeq 8.5 \times 10^{15} \textrm{GeV} (\al_{0.05})^{1/2}\widehat\lambda_3^{-1/2}$ which leads to a lower bound on the scale of inflation
$
\Lambda > 8.5 \times 10^{15} \textrm{GeV} (\al_{0.05})^{1/2}
$.

In brief, small field models can exhibit a variety of detectable phenomena: GW, running, and blue/red spectral index.
Any set of observables can be obtained from minimal potentials, hence reconstruction is impossible and identifying the inflaton will require additional knowledge about its interactions.

\begin{figure}[t]
{\psfig{figure=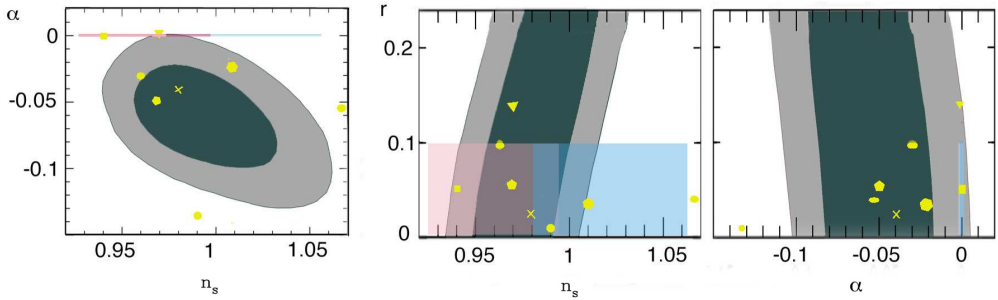,height=2.2in}}
\caption{\label{figure2} Model predictions for the eight models of Table~\ref{table:models} for various CMB observables on the background of the QUaD analysis of their CMB allowed region. The center and right panels show $r$ vs. $n_s$ and $\alpha=d n_s/d\ln k$ (respectively). The left panel shows $\alpha$ vs. $n_s$. The pink and blue rectangles depict the regions of parameter space that traditional models occupy.}
\end{figure}

\end{document}